 \documentclass[12pt]{article}   
 \usepackage{amsmath}
\usepackage{amsfonts}
\usepackage{graphicx} 

\newskip\humongous \humongous=0pt plus 1000pt minus 1000pt

\newif\ifdtup

\catcode`\@=11
 
\@addtoreset{equation}{section}
\def\theequation{\thesection\arabic{equation}}
 
\def\@normalsize{\@setsize\normalsize{15pt}\xiipt\@xiipt
\abovedisplayskip 14pt plus3pt minus3pt%
\belowdisplayskip \abovedisplayskip
\abovedisplayshortskip \z@ plus3pt%
\belowdisplayshortskip 7pt plus3.5pt minus0pt}
 
\def\small{\@setsize\small{13.6pt}\xipt\@xipt
\abovedisplayskip 13pt plus3pt minus3pt%
\belowdisplayskip \abovedisplayskip
\abovedisplayshortskip \z@ plus3pt%
\belowdisplayshortskip 7pt plus3.5pt minus0pt
\def\@listi{\parsep 4.5pt plus 2pt minus 1pt
     \itemsep \parsep
     \topsep 9pt plus 3pt minus 3pt}}
 
\relax

\catcode`@=12

\catcode`\@=11
 
\def\section{\@startsection{section}{1}{\z@}{3.5ex plus 1ex minus
   .2ex}{2.3ex plus .2ex}{\large\bf}}
 
\def\thesection{\arabic{section}.}

\def\appendix{\setcounter{section}{0}
 \def\thesection{Appendix \Alph{section}}
 \def\theequation{\Alph{section}.\arabic{equation}}}

\begin{document} 

\def\Ref#1{(\ref{#1})}
\def\be{\begin{equation}}
\def\ee{\end{equation}}  
\def\bea{\begin{align}}
\def\ena{\end{align}}
\def\ppn{\frac{2\pi}{N}}
\def\dl{{\dot\lambda}}
\def\ca{{\cal E}}
\def\field#1{{\mathbb #1}}
\def\ket#1{|{#1}\rangle}
\def\bra#1{\langle {#1} |}
\def\scalar#1#2{\langle{#1}|{#2}\rangle}
\font\msbm=msbm10
\renewcommand{\vec}[1]{\boldsymbol{#1}}

\def\non{\nonumber}
\def\De{\Delta}
\def\bqa{\begin{eqnarray}}
\def\eea{\end{eqnarray}}
\def\brc{\langle}
\def\ckt{\rangle}
\newcommand{\defi}{\stackrel{\rm def}{=}}
\def\de{\partial}
\def\si{\sigma}
\def\sb{{\bar \sigma}}
\def\Tr{\hbox{\rm Tr}}
\def\ker{\hbox{\rm ker}}
\def\dim{\hbox{\rm dim}}
\def\sup{\hbox{\rm sup}}
\def\inf{\hbox{\rm inf}}
\def\re{\hbox{\rm Re}}
\def\im{\hbox{\rm Im}}
\def\infi{\infty}
\def\nrm{\parallel}
\def\nrmi{\parallel_\infty}
\def\teo{\noindent{\bf Theorem}\ }
\def\all{\hbox{\rm all}}
\def\dirac{{\cal D}}
\def\om{\Omega}
\def\cm{{\rm cm}}
\def\erg{{\rm erg}}
\def\se{{\rm sec}}
\def\gram{{\rm gr}}
\def\mol{{\rm mol}}
\def\const{\hbox {\rm const.}}
\newtheorem{theorem}{Teorema}
\def\o{\over}
\def\bpm{\begin{pmatrix}}
\def\epm{\end{pmatrix}}
  \def\bmt{\begin{matrix}}
\def\emt{\end{matrix}}  
\def\bcs{\begin{cases}}
\def\ecs{\end{cases}}

\begin{titlepage}
{\hfill     IFUP-TH/23/2004} 
\bigskip
\bigskip

\begin{center}
{\large  {\bf  
On Cyclic Harmonic Oscillators
  } } 
\end{center}

\bigskip
\begin{center}
{\large  Kenichi KONISHI  and  Giampiero PAFFUTI
 }
\end{center}

\begin{center}

{\it  
Dipartimento di Fisica ``E. Fermi" -- Universit\`a di Pisa, \\
Istituto Nazionale di Fisica Nucleare -- Sezione di Pisa,  \\
  Via Buonarroti, 2,  Ed. C,   \\
  56127 Pisa, Italy\\ 
  e-mail: konishi@df.unipi.it, paffuti@df.unipi.it
   }
\end {center}
  
    \bigskip 
  \bigskip
  
  \noindent  
{\bf Abstract:}  
  
{ It is proven  that the energy of a  quantum mechanical harmonic oscillator with a generically time-dependent but  cyclic 
   frequency,  $\omega_{0}(t_{0})= \omega_{0}(0)$,   cannot decrease on the average  if the system is originally in a stationary state, after the system goes through a full cycle.  The energy exchange always takes place  in the direction from the macroscopic system (environment) to the quantum microscopic system.   }

\vfill  
 
\begin{flushright}
\today
\end{flushright}
\end{titlepage}
  
  \section {A Theorem  \label{sec:theorem} }
  
Many physical systems    reduce effectively  to that of a harmonic oscillator with a time-dependent frequency (we set $m=1$),  
  \be  H(t) =  {p^2 \o  2} +  { 1 \o 2}   \,  \omega(t)^2\, q^2,
  \label{HO}\ee 
  which is thus of certain interest \cite{DM}.    In particular,  the case of a periodic (cyclic)  variation
   of  $\omega$  such that 
  \be   \omega(t_0)= \omega(0)
  \ee 
  is of great interest. We shall not specify the time variation  of $\omega(t)$ otherwise. 
Suppose  furthermore that the system is initially  in one of the stationary states, $\ket  \psi$, with energy $E_{in}$.   We prove below,   independently
 of how the system goes  through the  cycle,  that 
\be  E_{fin}=  {\bar E}(t_{0}) =  \bra {\psi(t_{0})}  H  \ket {\psi(t_{0})}   \ge    E_{in}, 
\ee
{\it i.e.   on the average the microscopic system can never  give excess energy  to the external environment.  }
  Such a result might appear surprising at first sight,  since   during  a  generic time variation of external parameters, the system 
  can either  give  the energy away    to or absorb it  from the environment.   Also, such a theorem  certainly does not hold in general  in  a system with a finite number of independent states, such as a spin ${ 1\o 2}$ system in a varying magnetic field.   
      
   The proof is easiest  in the Heisenberg picture.  
   Heisenberg equations of motion  tell us that   $q(t)$, $p(t)$ are linear combinations  of $q$, $p$; they
   furthermore  preserve  the 
   commutation relation (we set $\hbar=1$)
   \be  [q(t), p(t)]=  i.  
   \ee
In  other words the time evolution is described by an  $Sp(2)$   transformation
\be   y(t)_{\alpha}=   S_{\alpha \beta}(t)  \,  y(0)_{\beta},
   \qquad      S_{\alpha \beta } S_{\gamma  \delta  }\varepsilon_{\beta \delta } = \varepsilon_{\alpha \gamma}\qquad S\varepsilon S^T = \varepsilon,   \label{sp2}   \ee
where we wrote $y_{\alpha} = (q, p)$.
 Or writing 
 \[ S = \bpm a & b\\ c & d  \epm,   \qquad  a,b,c,d \,\, {\rm real} \]
the condition  \Ref{sp2} reduces to 
\be ad - bc = \det(S)  = 1.  \label{eq3}\ee

The energy expectation value at time $t$  is given by
   \be {\bar E}(t) =  \bra {\psi}  H(t)   \ket {\psi} = { 1\o 2}    \bra {\psi}  p(t)^2 + \omega(t)^2 q(t)^2   \ket {\psi}. 
 \label{uptot}  \ee
 At the end of a cycle (we set $\omega(0)=1$),   
 \be E_{fin} = \frac{1}{2}\langle \psi| p_H(t_{0})^{2}
 + q_H(t_{0})^{2}  |\psi\rangle. 
\label{eq4}\ee
By using Eq.~(\ref{sp2})  and by defining 
\[ D = \frac{1}{2}( qp +pq) \]
the quadratic form  becomes 
\begin{eqnarray*}
&&(S_{11} q + S_{12} p)^2 + (S_{21}q + S_{22} p)^2 =\label{eq11a}   \\
&& = (S_{11}^2 + S_{21}^2) q^2 + (S_{12}^2 + S_{22}^2) p^2 +
2(S_{11} S_{12} + S_{21}S_{22}) D.   
\end{eqnarray*}
By using the Virial theorem 
\be \frac{1}{2}\langle \psi|p^2|\psi\rangle = 
\frac{1}{2}\langle \psi|q^2|\psi\rangle = \frac{1}{2} E_{in}
\ee
 and the fact  (valid for real wave functions) that 
\be
\qquad \langle \psi | D|\psi\rangle = 0 \ee
the final energy is given by 
\be E_{fin} = \frac{1}{2}E_{in}\left[S_{11}^2 + S_{21}^2 + S_{12}^2 + S_{22}^2\right].
\label{eq5}\ee
The problem is then  to find the minimum of a quadratic form
$  Q = a^2 + b^2 + c^2 + d^2   $
 under a constraint  $ ad-bc = 1$.
Upon introduction  of a Lagrange  multiplier,   the  extremum  of $a^2 + b^2 + c^2 + d^2 + 2\lambda(ad-bc - 1)$  
 is given by \footnote{As the set of the evolution matrices is unbounded, this extremum can only be a minimum. In an alternative proof given in (\ref{rsdeta32}) this fact is obvious. }
\[ \lambda = -1, \qquad  a = d, \qquad b = - c, \qquad S = \bpm a & b \\ -b & a  \epm, \qquad a^{2}+ b^{2}=1,   \]
that is,  when $S$ is orthogonal. In that case the  quadratic form takes the value $2$ and
therefore in general 
  \be E_{fin} = \frac{1}{2} \, E_{in} \, Q \geq E_{in}. \label{eq6}
\ee

 \subsection*{Remarks}

\begin{description}

\item{(i)}  In the adiabatic limit, the system ``follows'' the variation of the spectrum
while staying  in the  ``initial''  eigenstate, and comes back to the original state,  so we expect    $E_{f}\to E_{in}.$   In the sudden limit, the wave function does not make it to change as the external parameter goes through a (too) quick cyclic variation, so that   $E_{f} \to  E_{in}$ again.

\item{(ii)}
The above result is valid also  in the case of a forced oscillator. Consider
\be H = \frac{1}{2}(p^2 + \omega(t)^2 q^2) - \kappa(t) q,\label{eqa1}\ee
where  $ \kappa(t)$ is an arbitrary function with  $ \kappa(0) =  \kappa(T) = 0$.
The Heisenberg equations  are   
\begin{eqnarray}
\dot p &=& i [H,p] = - \omega^2 q +    \kappa  \non \\
\dot q &=& i[H,q] = p
\end{eqnarray}
so 
\be \ddot q = - \omega^2 q + {\dot  \kappa}.   \label{eqa2f}\ee 
Let us now consider a solution  $Q_c(t)$  of  \Ref{eqa2f} with the boundary condition
\[  Q_c(0) = \dot Q_c(0) = 0;  \]  
it follows immediately that $q$ is a sum of the homogeneous solution plus the particular one
 $Q_c$
\begin{eqnarray}
q_H(t) &=& Q_c(t) + q_H^{(\kappa=0)}(t),  \non  \\
p_H(t) &=& \dot Q_c(t) +p_H^{(\kappa=0)}(t).
\end{eqnarray}
One has then 
\[ E_f = E_f^{(\kappa=0)} + \frac{1}{2}\left( Q^2 + {\dot Q}^2\right)
+   ( Q_{c} \langle q_H^{(\kappa=0)}(t)  \rangle  +  {\dot Q}_{c} \langle p_H^{(\kappa=0)}(t)  \rangle  ).   \]
As the expectation values of  $q_H$ and  $p_H$ vanish in the initial stationary state it follows that 
\be E_f = E_f^{(\kappa=0)} + \frac{1}{2}\left( Q^2 + {\dot Q}^2\right) \ge  E_i, \label{eqa4}\ee
where the result of the preceding paragraphs has been used. 

\item {(iii)}  In perturbation theory the theorem can be easily seen to be valid.   By writing  $\omega(t)^{2} \, x^{2} =  \omega_{0}^{2}\, x^{2} + \delta \omega(t) \, x^{2}$,
$\delta \omega(t_{0})= \delta \omega(0)=0$,  the first-order transition probability 
\be  P_{f i } = { 1 \o \hbar^{2}} \left|\int_{0}^{t_{0}} dt \, {\delta \omega(t) \o 2 }\, e^{i \,\omega_{fi}\, t} \,( x^{2} )_{fi }
\right|^{2}
\ee
is larger 
 for the process 
$n \to n+2$ than  for $n \to n-2$,  hence 
 $\brc H \ckt \ge  E_{in}$. 

\item {(iv)}   In perturbation theory, one can actually verify  the theorem for a more general class of 
perturbing potentials. Indeed, it can be  easily seen  that for any perturbation of the form,
\be     \Delta V(t, x) = \delta \omega(t)  \, x^{N},  \qquad   \delta \omega(t_{0})= \delta \omega(0)=0,
\ee
the theorem is valid, as  $|(x^{N})_{n+m, n}| \ge   | (x^{N})_{n-m, n} |$  ($m>0$).

This would  suggest  that actually the theorem holds for a wider class of periodic potentials, $V(t_{0}, x)= V(0, x)$.

\item{(iv)}
The theorem obviously  applies for a  system of $N$ independent oscillators
\be  H= \sum_{i=1}^{N}  \,[\,   {p_{i}^2 \o  2} +  { 1 \o 2} \,    \omega_{i}(t)^2\, q_{i}^2 \,], 
\label{NOscil}\ee
with arbitrary, periodic variations of $\omega$,  $\omega_{i}(t_{0})= \omega_{i}(0), $  if the    
initial state is   in a stationary state $| n_{1}, n_{2}, \ldots, n_{N} \ckt.$

\item{(v)}   The theorem cannot hold for a generic initial pure state
 \footnote{We have however found a  {\it sufficient} condition for the theorem to be valid:  for initial states of the form $\psi= \sum_{n} a_{n }\psi_{n},$   $n= n_{0}+ 4 m$,   $m\in {\mathbb Z}$,  with a fixed $n_{0}$, the theorem can be shown to hold by a slight generalization of the proof given here. }.  Since  in some cases the final energy expectation value ({\it i.e.},  in the pure state $U(t_{0}) \, | \psi(0) \ckt  $) is strictly  higher than the initial value (Sec.~\ref{sec:example}), it suffices to consider the time-reversed process of such an evolution, to find a counter example \footnote{We thank Tomas Tyc for a useful communication on this point.}. 
	
On the other hand, for an initial state which is not an eigenstate of energy, the operational meaning 
of the theorem itself would become   somewhat unclear. 

\item{(vi)}  There is an  important case in which  the theorem applies for a initial  mixed state.  Consider the $N$ oscillator system of (\ref{NOscil}) and suppose that the system is   
originally   in the  thermal equilibrium with a heat bath  at temperature $T$.   The theorem then clearly applies in a statistical and quantum average sense.     

\item{(vii)}  The energy gain factor, $  R  \equiv   \langle\psi | H | \psi \rangle / E_{in}$,
is universal,  in the sense that it does not depend on the particular initial stationary state chosen. 

\item{(viii)}  Note that the (classical) canonical equations of motion 
and Heisenberg equations have the same form, and the evolution matrix $S$  are the same in both cases.  
As a result, an analogous theorem holds in  classical mechanics, if one takes an average over 
random initial values $(p,q)$  over a given (fixed-energy) trajectory. 

\end{description}  

\section {Example:  Inverse Linear Variation of the Frequency \label{sec:example}}

Consider  the oscillator Eq.(\ref{HO}) with  frequency varying   as  
\be  \omega(t) =   { \omega_{0} \o \lambda(t)}, \qquad   \lambda(t)= 1 + v \, t,  
\ee
where   $\omega_{0} $ adn $v$ are constants. 
From  Heisenberg equations   one gets  
\[  {d^{2} q\o d \lambda^{2}}  +   \Omega^{2}  { q \o  \lambda^{2}} =0, \qquad   \Omega=  { \omega_{0}  \o  v }.
\]
By setting $q(t) =  \lambda(t)^{\beta}$,  one gets a characteristic equation
\[  \beta \, (\beta -1) + \Omega^{2} =0, 
\]
with solutions
\be  \beta_{1,2} =   { 1\o 2} \pm \delta, \qquad  \delta \equiv  \sqrt{{1 \o 4}
- \Omega^{2}},   \ee
so that  the general solution of the  Heisenberg equations reads 
\be q_H = c_1 \lambda^{\beta_1} + c_2 \lambda^{\beta_2}, \qquad
 p_H =    v \, \left( c_1 \beta_1 \lambda^{\beta_1} +  c_2\beta_2 \lambda^{\beta_2}\right).
 \label{general} \ee
By imposing the initial condition one finds
 \begin{eqnarray}
 q_H(t) &=& \frac{1}{2 \,\delta }\left\{  
 \left(\beta_2 \, \lambda^{\beta_1} - \beta_1\, \lambda^{\beta_2} \right) \, q   +{ 1\o v}  \,
 \left(\lambda^{\beta_2}- \lambda^{\beta_1}\right)\,  p  \right\},   \non  \\
 p_H(t) &=& 
 \frac{v}{2 \,\delta \, \lambda }  \, \left\{
 \beta_1 \, \beta_2 \, \left(\lambda^{\beta_2}- \lambda^{\beta_1}\right)\, q  +  { 1\o v}   \left(\beta_2 \, \lambda^{\beta_2} - \beta_1 \, \lambda^{\beta_1}  \right)  \, p \right\}.      
 \label{Heisqp}\end{eqnarray}
As a check, consider the adiabatic limit, $v\to 0$, $\Omega \to \infty$.  One has  $\beta_{1,2} \simeq \pm i \, \omega_{0} /v$,
so that 
\[ \lambda^{\beta_{1,2}} \to  ( 1 + v \, t)^{\mp i \omega_{0} /v }  \to   e^{\mp i \, \omega_{0}  \,t},  
\]
and
\[  q_H(t) \to \frac{ q}{2} \, ( e^{-i \omega_{0}  t} + e^{ i \omega_{0}  t}) +
\frac{p}{2 i \omega_{0} }(-e^{-i\omega_{0}  t} + e^{i \omega_{0}  t}) = q\cos \omega_{0}  t + \frac{p}{ \omega_{0}}\sin \omega_{0}   t
\]
which is the correct result. 

Writing   Eqs.(\ref{Heisqp})  in the form of   Eq.(\ref{sp2})  with   $S=S(\omega_{0},v,\lambda) $ one gets,
by inserting this  in Eq.(\ref{uptot})  and  by using  the  Virial theorem,
\begin{eqnarray}
 \langle\psi | H | \psi \rangle &=&
\frac{1}{2}E_0\left[\frac{1}{\lambda^2}( S_{11}^2 + S_{12}^2) +  S_{21}^2 + S_{22}^2\right]=
\nonumber\\ &=&
\frac{E_0}{2}\frac{
\lambda^{2\delta} + \lambda^{-2\delta} + 2(4\delta^2-1)}{4\lambda  \delta^2}.    \label{eq10HO}
\end{eqnarray}
At time $t$,  $\lambda= 1 + v \, t,$   the   energy mean value (\ref{eq10HO}) can be larger or smaller than the original energy depending on  the sign of the velocity $v$ (hence whether the scale factor
$\lambda$ is smaller or larger than unity). 

However,  as we are most interested in  cyclic variations  of $\omega$, let us consider  the 
the evolution from the original frequency $\omega_{0}$  to a  final frequency $ \omega_{0} \o \lambda $, and then   back to  $\omega_{0}$.  The second half of the evolution 
 is simply described by the   transformation  $S=S({\omega_{0}\o \lambda}, - v, { 1 \o \lambda}) $
so that  the total evolution 
is  described by the Heisenberg evolution
\be   \left(\begin{array}{c}q_H(t_0) \\p_H(t_0)\end{array}\right) =S^{cyc} \,  \left(\begin{array}{c}q_H(0) \\p_H(0)\end{array}\right),
 \quad   S^{cyc} \equiv    S({\omega_{0}\o \lambda}, - v, { 1 \o \lambda}) \cdot
 S(\omega_{0},v,\lambda) 
\label{linearHE}\ee
where $S$ is defined by Eq.(\ref{Heisqp}).  

The results for more general frequency variations are  given in  \ref{sec:power} and  in   \ref{sec:expo}. 

\begin{figure}[httb*]
\begin{center}
\includegraphics[width=0.4\textwidth  ]{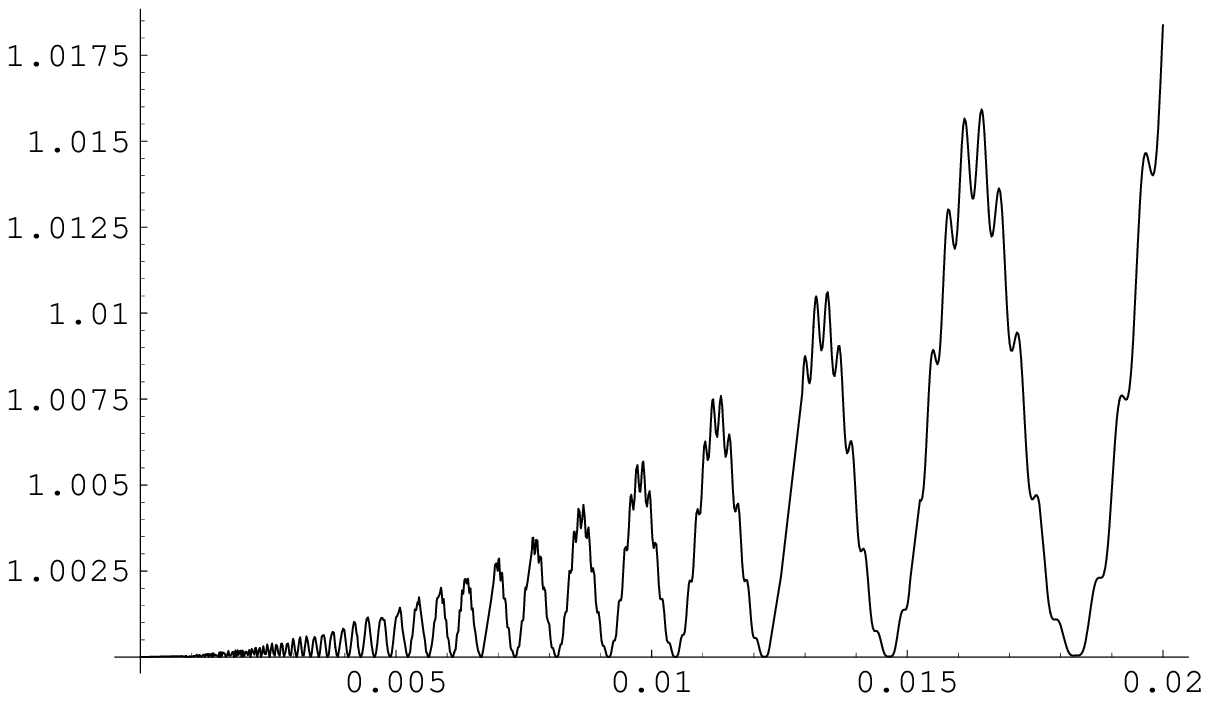} \hskip 0.1\textwidth \includegraphics[width=0.4\textwidth  ]{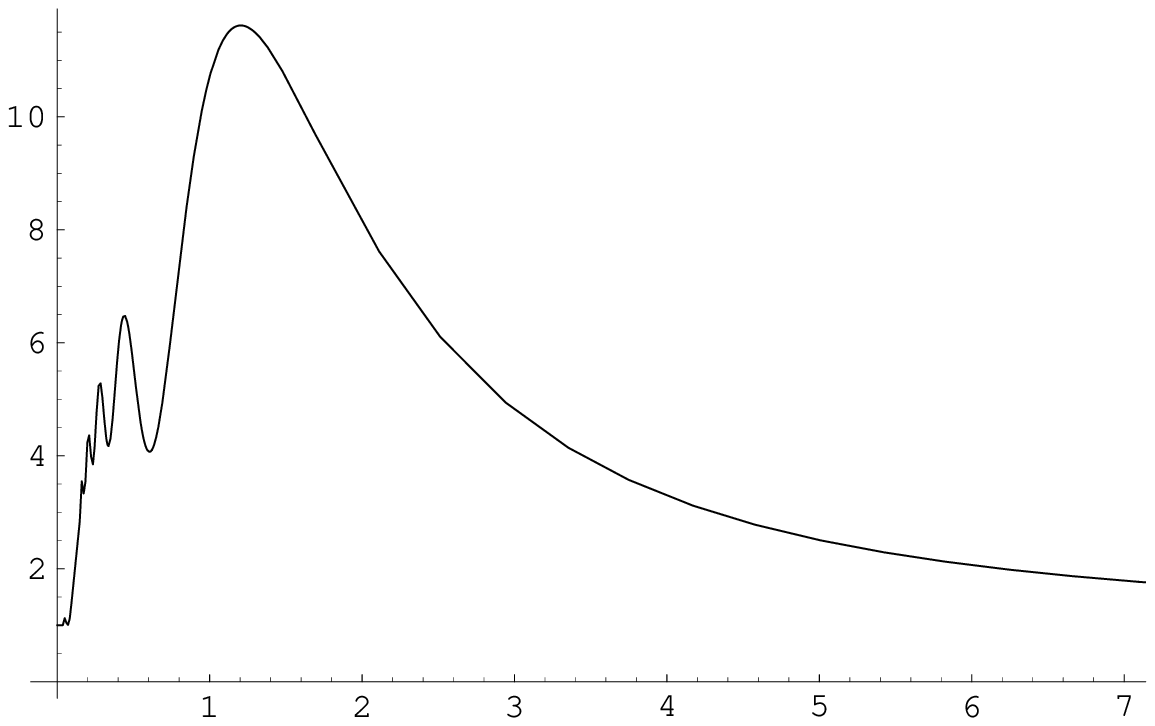} 
\caption{\footnotesize Energy gain factor  as a function of  $v$ for  fixed  $\lambda (= 10),$  at small and large $v$. }
\label{Energyres}
\end{center}
\end{figure}

In the case of the  linear variation (\ref{eq10HO}), (\ref{linearHE}), 
we have analysed  numerically  the energy gain factor 
$  R(\omega_{0},  v,  \lambda)=  \langle\psi | H | \psi \rangle / E_{0}=  { 1\o 2}\, \Tr \,[ \,S^{cyc} \,(S^{cyc})^{T} \,] $     for various values of   $\omega_{0},\,  v \, $ and $  \lambda $,   $\lambda$ being the 
scale factor  ($\omega = \omega_{0}/\lambda$)  at the moment of  the maximal  contraction ($\lambda <   1$) or expansion  ($\lambda >1$).   We find $  R(\omega_{0},  v,  \lambda) \ge 1$ always as expected,  but find also that: 

\begin{description}
\item  {(i)} At  fixed $\lambda$,  $  R(\omega_{0},  v,  \lambda) \to 1  $  both in the adiabatic ($v \to 0$) and  impulse approximation
($v \to  \infty $)  limits, as expected; 
\item   {(ii)}  There are various resonance  effects at small $v$  (see Fig.~\ref{Energyres});  in particular, as a function of the velocity $v$ at a fixed 
$\lambda$,  $R$  reaches a maximum of order of  $O(1/\lambda)$   (for  $\lambda <   1$)  or  
 $O(\lambda)$   (for  $\lambda >  1$),   and then rather smoothly approaches  the impulse-approximation    value $1$ asymptotically;  
 \item {(iii)} When the system goes through $N$ cycles, the maximal  energy gain factor  (which occurs at certain  critical velocity)  behaves as 
 $R \propto   (1/\lambda)^{N}$  or  $R \propto  \lambda^{N}$, a huge factor if $\lambda $ is large; 
\item{(iv)} There are values of ($v, \lambda$) at which $R$  attains values either exactly equal to or very close to unity (Fig.~\ref{Energyres}). 
  \item{(v)} As a function of $\lambda$, the maximum of $R$ grows indefinitely as   $\lambda\to 0$ or as  $\lambda\to \infty. $
\end{description}
  
\section{Planck Distribution inside an Oscillating Cavity}

As a possible physical  application of our considerations, let us consider the black body spectrum confined in a perfectly reflecting three-dimensional box  of linear dimension $L$.  The electromagnetic fields are described by the Lagrangian  
\[  L = {1 \over 8 \pi} ({ L\over 2})^3 [\sum_{\bf n} {1\over c^2}
 ({\dot Q_{1{\bf n}}}^2 +{\dot Q_{2{\bf n}}}^2)
 - \sum_{\bf n}  {\bf k}_{\bf n}^2 ( Q_{1{\bf n}}^2 + Q_{2{\bf n}}^2)],
\] 
where  $ {\bf k}_{\bf n} \equiv \pi {\bf n}/L.$  
A redefinition of $Q_{i{\bf n}}$ and a Legendre transformation leads to the Hamiltonian per unit volume
\[  H=\sum _{\bf k}  \left({c^2 \over 4 } {\bf p}_{(1)}^2  + {\bf k}^2  
{\bf q}_{(1)}^2\right)  + \sum _{\bf k} \left({c^2 \over 4 } {\bf p}_{(2)}^2  + 
{\bf k}^2   {\bf q}_{(2)}^2\right).
 \label{Helmag}\]
At temperature $T$  the energy distribution is 
\be   u(\nu)\,d\nu  =  {8\pi  \over c^3}{  h \nu \over e^{ h \nu /kT} -1}
\nu^2\, d\nu.  \label{Planck} \ee

Suppose that at certain moment the box starts to contract or expand  with a constant linear velocity.  How does the energy distribution change with the linear size of the box? 

At time $t=0$ various modes are distributed according to the Planck distribution,
 (\ref{Planck}).   Each mode simply transforms as  in Eq.(\ref{Heisqp}): 
($ |\delta| \gg 1,\,\,$ $\delta= i |\delta|$),
\be  E_n \to  \brc H \ckt =   \frac{E_n}{2}\frac{
\lambda^{2\delta} + \lambda^{-2\delta}  +  
2(4\delta^2-1)}{4\lambda  \delta^2}\simeq   { E_n \o   \lambda }   \label{rsb2eq10}
\ee
 Since
\[ \omega =  { \pi |{\bf n}| c \o  L_0}, \qquad   L(t)=  L_0 (1 + v  \, t),  \quad  |V| =   L_0\, |v|  \ll   c,  
\]
\be   \Omega =  { \omega \o v}=  {\pi |{\bf n}| c \o |V| } \gg 1, \qquad  \delta = \sqrt{ { 1\o 4}  - \Omega^2} \simeq i  {\pi |{\bf n}| c \o |V| }  
\ee
the process is adiabatic for all modes.  The distribution remains Planckian; each mode is either red-shifted (in an expanding box) or shifted towards ultraviolet (a contracting box), simply by the scale factor $\lambda$.

Actually, the system is not so interesting as an application  of the  general theorem mentioned in Section 1,  as it is entirely adiabatic;
the only reason we discuss it here is its  possible relevance to the so-called sonoluminescence phenomenon. 

In the  single-bubble sonoluminescence \cite{SL}, a gas bubble trapped in a liquid is made oscillating radially  by acoustic  waves, and at the moments its radius attains the minimum the bubble emits  pulse of light in the visible to ultraviolet wavelength range. 
In a typical experiment, the linear dimension of the bubble contracts up to a factor of $\lambda = 10^{-4}$:  
this is compatible with the observed increase of an  effective temperature of a factor $10^4$ or more.
The energy density increases according to the Stefan's law, 
 \be U = \sigma T^4; \quad \sigma = 7.64 \cdot 10^{-15 } \erg \cm^{-3}
 {\hbox {\rm K}}^{-4} \label{Stefan}  \ee 
 but since the volume itself decreases by a factor $\lambda^3 $ the total energy of blackbody radiation increases 
 only by a factor $ { 1\o \lambda}$.  It is possible that this  excess energy is released  in the form of visible light pulse at each cycle \footnote{In an experiment at Laurence Livermore Ntional Laboratory \cite{SL}  
 $10^5 - 10^6$  photons are observed within a pulse, with the energy of order of $ 10^{-7 }  \, \erg $  which seems to be compatible with such a rough estimate. Note also that the visible lights correspond to the effective temperature of
 $T=10^4 -10^5$   or to  the energy for a single photon of order of  
$   h \nu \sim k T  \sim    10^{-12} \, \erg. 
$}.

\section{Conclusion}

A harmonic oscillator with a time-dependent frequency (or an ensemble of such oscillators)   experiences  a full cycle,  with the system originally in a stationary state.  After the cycle,  the energy expectation value is predicted never to decrease, independently of the way the parameters vary with time during the cycle. Our result is somewhat reminiscent of the law of entropy increase, but concerns  the evolution of  quantum mechanical pure states,  and no information loss is implied.   A generic  non-adiabatic ``disturbance''  by the external force always does  work on  the microscopic system on the average, increasing its energy. The energy flow is always in the direction from the macroscopic system to the microscopic one.  In  other words, the quantum mechanical  harmonic oscillator cannot act as a perpetual machine, nor  produce a net energy gain. 
 A spaceship cannot continue her 
journey forever, getting its energy supply  from  the inexhaustible zero-point energy of the vacuum \cite{LN}.

\section*{Acknowledgment}

The authors  are indebted to David M. Brink for many enlightening discussions (in particular on the applicability of the theorem in perturbation theory and in classical mechanics),  and for sharing their  enthusiasm.  Thanks are also due to Francesco Maccarrone for discussions  on the sonoluminescence. Numerical analyses have been done by use of Mathematica (Wolfram Research) and MatLab (MathWorks).

\appendix

\section{General Power Dependent Frequencies \label{sec:power} }

\begin{figure}[httb*]
\begin{center}
\includegraphics{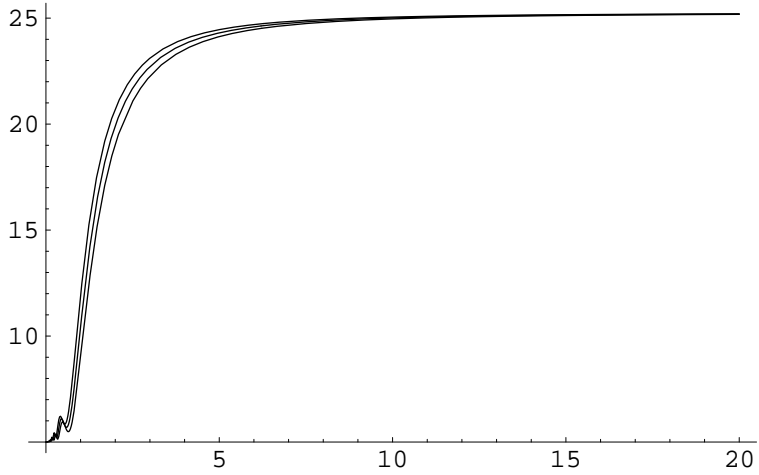}  
\caption{\footnotesize Energy as a function of  $v$ for a fixed  $\lambda = 1/10$,  with $k=-2,-3,-4$.
Expected asymptotic value is $1/4(1+100) = 25.25$.}
\label{Energyas}
\end{center}
\end{figure}

Consider  a case of a generic power-behaved frequency,   
\be H(t) = \frac{1}{2m}p^2 + \frac{1}{2} \, m \, \omega_{0}^2 \, z^{k-2} \, x^{2}, 
\qquad z = 1 + v t. 
\label{rsb2eq11}\ee
For $k=0$ we recover the case discussed in the text.   In the following, we shall set $m=1,\omega_{0} = 1.$
Note that  $v$ has the dimension of a frequency, so that to recover the dependence on $\omega_{0}$  it suffices to replace 
it by $v/\omega_{0}.$
The Heisenberg equation of motion gives   
\[ \ddot q_H + z^{k-2} q_H = 0.   \label{rsb2eq12}\]
Multiplying this by  $1/v^2$ one gets 
\be \frac{d^2}{d z^2}q_H + {1 \o v^2}   z^{k-2} q_H = 0. \label{rsb2eq13}\ee 
Its general solution is of the form,  
\be q_H(t) = A \sqrt{z} J_{1/k}(\frac{2}{kv} z^{\frac{k}{2}}) +
B \sqrt{z} Y_{1/k}(\frac{2}{kv} z^{\frac{k}{2}})\label{rsb2eq14}\ee 
where  $J,Y$  are the  Bessel functions of the first and second kind, respectively. 
Differentiation with respect to time yields  
\[ p_H(t) = v \frac{d}{d z} q_H(t).  \label{rsb2eq15}\]
The coefficients $A$ and $B$ are determined by imposing the initial conditions
\[ \left. q_H\right|_{z=1} = q\qquad \left. p_H\right|_{z=1} = p: \label{rsb2eq16}\]
\begin{eqnarray}
A &=& -\frac{\pi}{ k v}\left[ q\cdot( Y_{1+\frac{1}{k}}(\frac{2}{kv}) + (p - v q)\cdot Y_{\frac{1}{k}}(\frac{2}{kv})\right];  \non \\
B &=& \frac{\pi}{ k v}\left[ q\cdot( J_{1+\frac{1}{k}}(\frac{2}{kv}) + (p - v q)\cdot J_{\frac{1}{k}}(\frac{2}{kv})\right]. 
\end{eqnarray}
The expression for 
\be E(t) = \langle 0| \frac{1}{2}\,  p_{H}(t)^2 + \frac{1}{2}z^{k-2} q_{H}(t)^2 |0\rangle. \label{rsb2eq17}\ee
 is quite complicated but can be analyzed numerically.  We note that: 
\begin{itemize}
\item [1)] The case $k=-2$ reduces to the known result; 
\item [2)] The behavior is qualitatively similar for all  $k$,   with a phase of monotonous increase following  the initial, oscillating phase, as a function of $v$ (see Fig.~\ref{Energyas});  
\item [3)] At fixed  final ``physical scale'', 
$ z_{fin} = \lambda^{-\frac{2}{k-2}}, \quad  z_{fin}^{k-2} = \lambda^{-2}, $
the asymptotic  behavior in $v$  turns out to be always  the same as in the case  $k=-2$, {\it i.e., } 
\be E_{f} \simeq  \frac{\omega}{4}(1 + \frac{1}{\lambda^2}).  \label{asymptotic}\ee
This  assertion is based on a graphical evidence (see Fig.~\ref{Energyas}), for  the moment \footnote{For generic order, the functions $J,Y$ have essential singularities at $v=\infty$, which prevent us from
analyzing  either adiabatic or  large $v$ limit analytically. }.
 
\end{itemize}

\section{Exponential Dependence \label{sec:expo}}

As a second example, let us consider  the Hamiltonian, 
\be H = \frac{1}{2} \, p^2 + \frac{1}{2} \, e^{2 v t} \,q^2. \ee
The Heisenberg equation is
\[ \ddot q_H + e^{2 vt } q_H = 0. \]
Changing the variable to  $\tau = vt$ one has 
\[ \frac{d}{d \tau^2}q_H + \frac{1}{v^2}e^{2 vt } q_H = 0\]
which has 
\begin{eqnarray} 
q_H(t) &=& A\, J_0(\frac{1}{v} e^{vt}) + B \,Y_0(\frac{1}{v} e^{vt})\label{rsb2eq18}\non \\
p_H(t) &=& \frac{d q_H}{dt} = -A \,e^{vt}J_1(\frac{1}{v} e^{vt}) - B\,   e^{vt} \, Y_1(\frac{1}{v} e^{vt})
\end{eqnarray}
as general solutions.  As before the coefficients are determined by the initial condition:
\[
A = -\frac{\pi}{2v}\left[ p\, Y_0(\frac{1}{v}) + q\, Y_1(\frac{1}{v})\right], \qquad  
B= \frac{\pi}{2v}\left[ p\, J_0(\frac{1}{v}) + q \, J_1(\frac{1}{v})\right]. 
\]

The mean energy can be computed as in the preceding cases. 
\begin{eqnarray} E(t) &=& \frac{\pi^2 z^2}{16\pi^2}\left\{-2 J_0(\frac{z}{v})Y_0(\frac{z}{v})\left[
J_0(\frac{1}{v})Y_0(\frac{1}{v})+J_1(\frac{1}{v})Y_1(\frac{1}{v})\right] \right. \non\\
&& + J_0^2(\frac{z}{v})\left[Y_0^2(\frac{1}{v}) + Y_1^2(\frac{1}{v})\right] +
J_1^2(\frac{z}{v})\left[Y_0^2(\frac{1}{v}) + Y_1^2(\frac{1}{v})\right] \non \\
&& +2 J_1(\frac{z}{v})Y_1(\frac{z}{v})\left[
J_0(\frac{1}{v})Y_0(\frac{1}{v})+J_1(\frac{1}{v})Y_1(\frac{1}{v})\right]+  \non \\
&&+ \left.\left[J_0^2(\frac{1}{v}) + J_1^2(\frac{1}{v})\right]\left[Y_0^2(\frac{z}{v})+Y_1^2(\frac{z}{v})\right]\right\}
\end{eqnarray}
where  $z = e^{vt}$. 
It is also possible to get the asymptotic behavior in 
 $v$ at fixed   $z$: 
\be E_{as} = \frac{1}{4}(1 + z^2) - \frac{1}{16 v^2} (z^2-1)^2,  \label{rsb2eq19}\ee
which is compatible  with (\ref{asymptotic}) as  $z = 1/\lambda$.

\section {Creation and Annihilation Operators} 

The whole problem can be  analyzed by use of creation and annihilation operators.  We introduce at each instance the variables $q_{i}(t),\, p_{i}(t)$   in which  the frequency is
diagonal, $\omega_{ij}(t)=\delta_{ij} \omega_{i}(t)$; then define  $a_{i}(t),a_{i}^\dagger(t)$ in the standard manner in terms of $q_{i}(t),\, p_{i}(t)$. 
The time evolution introduces a linear transformation among  $a_{i}(t),a_{i}^\dagger(t)$, 
which has a  general form,  
\[  a^\dagger_i \to  A_{ik}a^\dagger_k + B_{ik}a_k,       \qquad 
a_i   \to A^*_{ik}a_k + B^*_{ik}a^\dagger_k. \] 
The coefficients must be such that  the canonical commutation relations
are preserved  (in a matrix form):
\be
A A^\dagger - B B^\dagger = 1, \qquad  \label{rsdeta31}\\
A B^T - B A^T = 0.  \ee
For a single oscillator, the theorem of Section~\ref{sec:theorem}  can be immediately proven: 
\begin{eqnarray}
E_f &=& \frac{\omega}{2} + \omega\, \langle n| (a')^\dagger a' |n\rangle =
\frac{\omega}{2}+\omega  \,\langle n| (|A|^2 + |B|^2)a^\dagger a |n\rangle =\nonumber\\
&=& \frac{\omega}{2} + n\, (1 + 2 |B|^2) \, \omega\ge  \omega\, (n + {1\o 2} ) = E_{in}. 
\label{rsdeta32}
\end{eqnarray}
         
  For $N$ oscillators which are originally in a stationary state  (ignoring the zero point energy), 
\begin{eqnarray*}
E_i &=& \langle \Psi| \sum_i \omega_ia^\dagger_ia_i|\Psi\rangle =
\sum_i \omega_i n_i \\
E_f &=& \langle \Psi| \sum_i \omega_i{a'}^\dagger_ia'_i|\Psi\rangle = 
\langle \Psi|\sum_i \omega_i 
(A_{ik}a^\dagger_k + B_{ik}a_k)(A^*_{i\ell}a_\ell + B^*_{i\ell}a^\dagger_\ell)
|\Psi\rangle =\\
&=& \sum_{i,k} \omega_i (A_{ik}A^*_{ik} n_k + B_{ik}B^*_{ik}(n_k+1)). 
\end{eqnarray*}
For the diagonal matrices $A$ e $B$  (uncoupled oscillators), (\ref{rsdeta32})  trivially generalizes. 
For coupled oscillators, the generalization is not obvious. 
  
We content ourselves  here  with a somewhat weaker  result:   it states that the occupation number (number of phonons)
does not diminish under  cyclic variations of frequency.    Indeed, 
\begin{eqnarray*}
N_i &=& \langle \Psi| \sum_i a^\dagger_ia_i|\Psi\rangle =
\sum_i  n_i \\
N_f &=& \langle \Psi| \sum_i {a'}^\dagger_ia'_i|\Psi\rangle = 
\langle \Psi|\sum_i  
(A_{ik}a^\dagger_k + B_{ik}a_k)(A^*_{i\ell}a_\ell + B^*_{i\ell}a^\dagger_\ell)
|\Psi\rangle =\\
&=& \sum_{i,k}  (A_{ik}A^*_{ik} n_k + B_{ik}B^*_{ik}(n_k+1))
\end{eqnarray*}
Eq.\Ref{rsdeta31} implies
\[ \sum_i A_{ik}A^*_{ik} = \sum_i B_{ik}B^*_{ik} + 1\]
therefore
\be N_f = \sum_k n_k + \sum_{i,k}B_{ik}B^*_{ik}(2 n_k+1) \ge N_i.  
\label{rsdeta35}\ee

\end{document}